\RequirePackage{amsmath}
\documentclass [twocolumn]{svjour3}
\usepackage[english]{babel}
\usepackage{booktabs}
\usepackage[numbers,sort]{natbib}
\usepackage{bm}
\usepackage{graphicx}
\usepackage[utf8]{inputenc}
\usepackage[T1]{fontenc}
\usepackage[switch]{lineno} 
\usepackage[hidelinks]{hyperref}

\begin{document}

\bibliographystyle{plainnat}

\title{Rucio -- Scientific data management}

\author{Martin Barisits \and Thomas Beermann \and Frank Berghaus \and Brian Bockelman \and Joaquin Bogado \and David Cameron \and Dimitrios Christidis \and Diego Ciangottini \and Gancho Dimitrov \and Markus Elsing \and Vincent Garonne \and Alessandro di Girolamo \and Luc Goossens \and Wen Guan \and Jaroslav Guenther \and Tomas Javurek \and Dietmar Kuhn \and Mario Lassnig \and Fernando Lopez \and Nicolo Magini \and Angelos Molfetas \and Armin Nairz \and Farid Ould-Saada \and Stefan Prenner \and Cedric Serfon \and Graeme Stewart \and Eric Vaandering \and Petya Vasileva \and Ralph Vigne \and Tobias Wegner}

\institute{
\begin{tabbing}
E-Mail:~~~~~~~\={\em rucio-dev@cern.ch}\\
Web: \>{\em https://rucio.cern.ch}\\
Repository: \>{\em https://github.com/rucio}
\end{tabbing}}

\date{Received: 1970-01-01 / Accepted: 1970-01-01}

\maketitle
\begin{abstract}
Rucio is an open-source software framework that provides scientific collaborations with the functionality to organize, manage, and access their data at scale. The data can be distributed across heterogeneous data centers at widely distributed locations. Rucio was originally developed to meet the requirements of the high-energy physics experiment ATLAS, and now is continuously extended to support the LHC experiments and other diverse scientific communities. In this article, we detail the fundamental concepts of Rucio, describe the architecture along with implementation details, and report operational experience from production usage.
\end{abstract}

\setcounter{tocdepth}{2}


\section{Introduction}

\subsection{Overview}

For many scientific projects, data management is becoming an increasingly complex and complicated challenge. The number of data-intensive instruments generating unprecedented volumes of data is growing and their surrounding workflows are becoming more complex. Their storage and computing resources are heterogeneous and can be distributed at numerous geographical locations belonging to different administrative domains and organizations. These locations do not necessarily coincide with the places where data is produced nor where data is stored, analyzed by researchers, or archived for safe long-term storage. The ATLAS Experiment~\citep{atlas:paper,atlas:website} at the Large Hadron Collider (LHC)~\citep{lhc:paper} at CERN~\citep{cern:paper} has such conditions. To fulfill the needs of the experiment, the data management system Rucio has been developed to allow ATLAS to manage its large volumes of data in an efficient and scalable way. Existing data handling systems focused on single tasks, e.g., transferring files between data centers, synchronizing directory contents across the network, dataflow monitoring, or reports of data usage. Rucio has been built as a comprehensive solution for data organization, management, and access for scientific experiments which incorporates the existing tools and makes it easy to interact with them. One of the guiding principles of Rucio is dataflow autonomy and automation, and its design is geared towards that goal. Rucio is built on the experiences of its predecessor system DQ2~\citep{dq2} and modern technologies, and expands on functionality, scalability, robustness, and efficiency which are required for data-intensive sciences. Within ATLAS, Rucio is responsible for detector data, simulation data, as well as user data, and provides a unified interface across heterogeneous storage and network infrastructures. Rucio also offers advanced features such as data recovery or adaptive replication, and is frequently extended to support LHC experiments and other diverse scientific communities. 

In this article, we describe the Rucio data management system. We start by detailing the requirements of the ATLAS experiment and the motivation for Rucio. In Section \ref{sec:concepts} we describe the core concepts and in Section \ref{sec:architecture} the architecture of the system, including the implementation highlights. Section \ref{sec:functionality} explains how the concepts and architecture together are used to provide data management functionality. We continue in Section \ref{sec:ops} with a view on the operational experience with a focus on deployment, configuration, and system performance, and in Section \ref{sec:advanced} with details on advanced workflows that can be enabled through Rucio. We close the article in Section \ref{sec:summary} with a summary, an overview of the Rucio community, and outlook on future work and challenges to prepare Rucio for the next generation of data-intensive experiments.

\subsection{ATLAS Distributed Computing}

ATLAS is one of the four major experiments at the Large Hadron Collider at CERN. It is a general-purpose particle physics experiment run by an international collaboration and it is designed to exploit the full discovery potential and the huge range of physics opportunities that the LHC provides. The experiment tracks and identifies particles to investigate a wide range of physics topics, from the study of the Higgs boson~\citep{higgs} to the search for supersymmetry~\citep{susy}, extra dimensions~\citep{extra}, or potential particles that make up dark matter~\citep{darkmatter}. The physics program of ATLAS is thus very diverse and requires a flexible data management system.

ATLAS Distributed Computing (ADC)~\citep{atlas:adc} covers all aspects of the computing systems for the experiment, across more than 130 computing centers worldwide. Within ADC, Rucio has been developed as the principal Distributed Data Management system, integrating with essentially every other component of the distributed computing infrastructure, most importantly the workflow management system PanDA~\citep{panda} and the task definition and control system ProdSys~\citep{prodsys}. ADC also does diverse research and development projects on databases, analytics, or monitoring. ADC is also in charge of the experiment computing operations and user support~\citep{atlas:tierz,atlas:tdaq,atlas:software}, which takes care of the needs of the physics community, and is in charge of commissioning and deployment of the computing services. Rucio is well-embedded into the work environments of ADC using both Agile and DevOps methodologies~\citep{agile}.

One of the most critical aspects of Rucio within ADC is the smooth interaction with the workflow management systems PanDA and ProdSys. In general, users do not see the data management system when they submit their simulation and analysis jobs, during the execution of the jobs, and after the jobs have finished. PanDA instructs Rucio that a job needs input files at a particular data center and Rucio will ensure that the files are made available at the given destination, and will also ensure that newly created files during job execution are promptly registered. In case of competing requests on constrained network or storage resources Rucio will schedule the dataflow to ensure fair usage of the available resources. Only at the very last stage, physicists might use Rucio directly to download job output files from globally distributed storage to their laptops or desktop nodes.

\subsection{Storage, network, and transfer providers}

The distributed computing infrastructure used by ATLAS comprises several systems which have been developed independently, through common research initiatives, or through high-energy physics focused projects. Most importantly, this includes different types of storage systems and their network connections.

The majority of storage systems in use are EOS~\citep{eos}, dCache~\citep{dcache}, XrootD~\citep{xrootd}, StoRM~\citep{storm}, Disk Pool Manager (DPM)~\citep{dpm}, and Dynafed~\citep{dynafed}. The dCache and CASTOR~\citep{castor} systems are used for tape storage. All these storage systems provide access to the data via several protocols, most importantly gsiftp~\citep{gsiftp}, SRM~\citep{srm}, ROOT~\citep{root}, WebDAV~\citep{webdav}, and S3~\citep{s3}. All except S3 can be used for direct storage-to-transfer transfers. The storage backends are commonly pools of disks, shared file systems, or magnetic tape libraries. The storage systems also handle authentication and authorization in various ways, such as X.509~\citep{x509} certificates or access control lists. Rucio is able to interact with these storage systems directly and transparently via custom implementations of the access protocols. Direct read access to magnetic tape libraries is also supported, however the clients will have to wait for the tape robot to stage the file if it is not in the tape buffer. Writing to tapes is supported via an asynchronous mechanism, to ensure efficient packing of files on the magnetic bands. If new storage systems enter the landscape, Rucio will be able to interact with them automatically if they support any of the existing protocols. Alternatively, new protocols can be implemented with plugins following the Rucio-internal interaction interfaces which mimic common POSIX operations such as {\em mkdir} or {\em stat}.

The network infrastructure is provided by multiple National Research and Education Networks (NRENs), most importantly ESnet~\citep{esnet}, Geant~\citep{geant}, Internet2~\citep{internet2}, SURFnet~\citep{surfnet}, and NORDUnet~\citep{nordunet}. The LHCOPN~\citep{lhcopn} connects the larger data centres directly, whereas LHCONE~\citep{lhcone} is a virtual network overlay across multiple NRENs. The smaller institutes are commonly connected via 40 Gbps links, the larger ones with 100 Gbps links. In total, the Worldwide LHC Computing Grid (WLCG) provides 3 Tbps network capacity across all available links~\citep{wlcgnetcapacity}. Commercial cloud storage providers have peering points for improved throughput from the NRENs into their private networks. Traffic can also be routed over the commodity internet as a fallback. In general, Rucio does not see the underlying network infrastructure, however, it can use historical network metrics such as throughput, packet loss, or latency to select better data transfer paths.

The middleware to establish direct {\em storage to storage} transfers over the network, commonly called {\em third party copy}, is provided by the {\em File Transfer System} (FTS)~\citep{fts}. FTS is a hard dependency for Rucio instances which require third party copy. FTS establishes connections between storage systems using the required protocols and ensures that the files are correctly transferred over the networks. Rucio decides which files to move, groups them in transfer requests, submits the transfer requests to FTS, monitors the progress of the transfers, retries in case of errors, and notifies the clients upon completion. If there are multiple FTS servers available, Rucio is able to orchestrate transfers among them for improved parallelism and reliability.

\section{Concepts}
\label{sec:concepts}

\subsection{Overview}

The main concepts supported by Rucio cover {\em namespace}, {\em accounts}, {\em storage}, {\em subscriptions} and {\em replication rules}. The {\em namespace} is responsible for addressing the data in various ways, {\em accounts} handle authorization, authentication, and permissions, {\em storage} provides a unified interface to the distributed data centers, {\em subscriptions} are used for large-scale dataflow policies, and {\em replication rules} ensure the consistent distributed state of the namespace on the storage.

\subsection{Namespace}

\begin{figure}[t]
  \includegraphics[width=\linewidth]{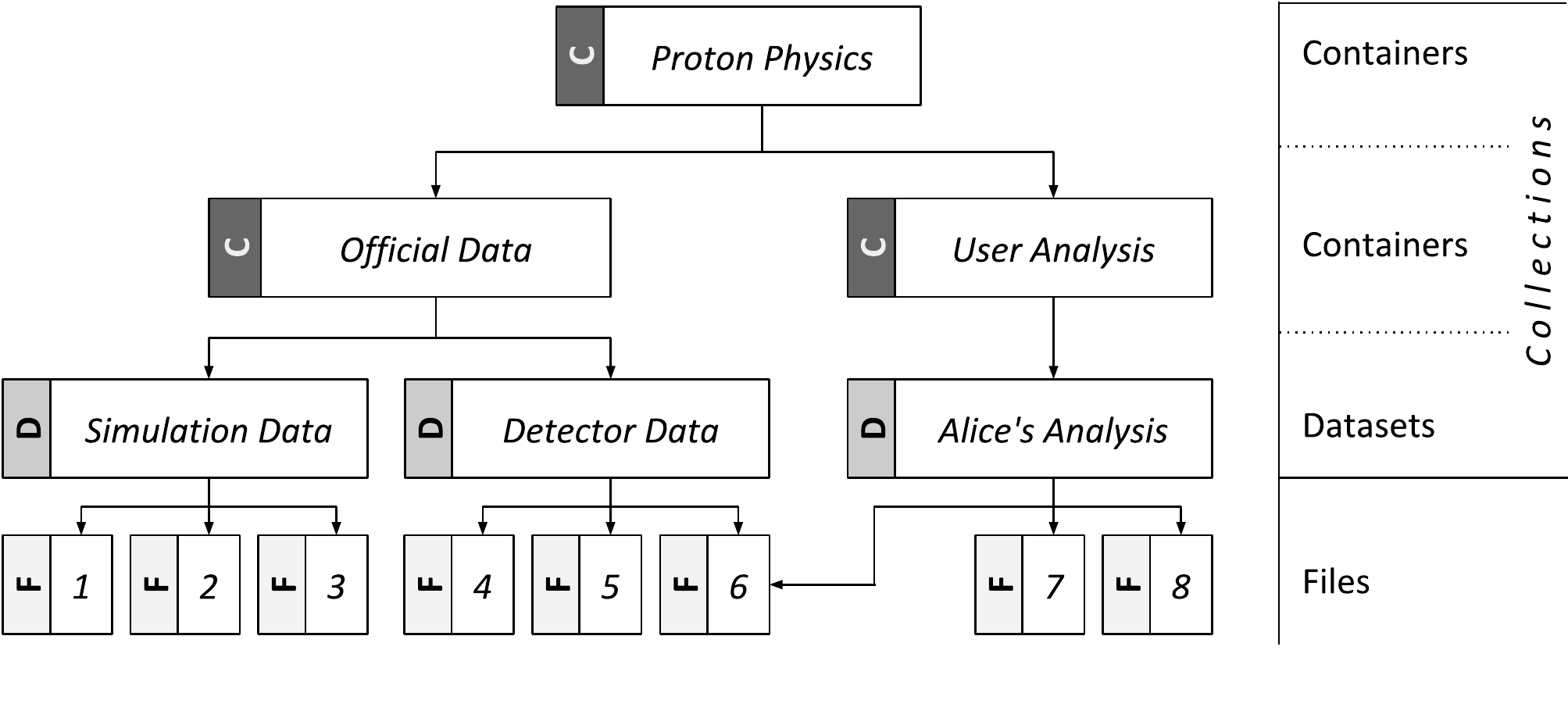}
  \caption{The namespace is organized with collections and files. Collections can either be containers or datasets. Containers consist of containers or datasets. Datasets consist of files only. Files can be in multiple datasets.}
  \label{fig:cont_ds_file_overview}
\end{figure}

Data in Rucio is organized using Data Identifiers (DIDs). There are three levels of granularity for DIDs:  files, datasets, and containers. The smallest unit of operation in Rucio is the file, which corresponds to the actual files persisted on storage. Datasets are a logical unit which are used to group sets of files to facilitate bulk operations on them, e.g., in transfers and deletions, but also for organizational purposes. The files in a dataset do not necessarily need to all be placed at the same storage location but can be distributed over multiple data centers as distributed datasets. Containers are used to group datasets and also to organize large scale groupings, such as annual detector data output or collections of simulation production with similar properties. Datasets and containers are referred to as collections. As shown in Figure \ref{fig:cont_ds_file_overview} this allows for multi-level hierarchies of DIDs, as DIDs can be overlapping. In this example, a proton physics experiment is split into curated {\em Simulation Data} and {\em Detector Data}, and the {\em User Analysis} data. Alice has created a research dataset named {\em Alice's Analysis} that contains the {\em F6} data from the detector which is necessary for her computation, which eventually produced two output files {\em F7} and {\em F8}.

All DIDs follow an identical naming scheme which is composed of two strings in a tuple: the {\em scope} and the {\em name}. The combination of scope and name must be unique, and is denoted via colons, e.g., the unique DID {\em data2018:mysusysearch01} is part of scope {\em data2018}. The scope thus partitions the global namespace. At least a single scope must exist, however the use of multiple scopes can be beneficial to data organisation. Straightforward use cases for multiple scopes are to easily separate instrumentation data from simulation data from user-created data, or to allow fine-grained permissions.

DIDs are identified forever. This implies that a DID, once used, can never be reused to refer to anything else at all, not even if the data it referred to has been deleted from the system. This is a critical design decision when dealing with scientific data, otherwise it would be possible to modify or exchange data used in previous analyses without warning. This does not mean that scientific data is required to be immutable, only that Rucio enforces a name change when data has been changed.

Rucio also supports a standardized naming convention for DIDs and can enforce this with a schema. This includes limits on overall character length, e.g., to reflect file system limitations, and that the names could be composed of fields referencing high level metadata such as the file format and processing version identifiers, as well as other metadata which is useful to connect DIDs with experiment operations. As an example, in ATLAS real event data DIDs contain the year of data taking and the ATLAS run number, and simulated event DIDs contain an identifier specific to the fundamental physics process being simulated. One important built-in metadata are the file checksums, which are rigidly enforced by Rucio whenever any file is accessed or transferred. The two checksum algorithms MD5 and Adler32 are supported. Experiment-internal metadata can also be enforced to follow a certain schema or uniqueness, such as the globally unique identifiers (GUIDs) used by ATLAS. All metadata is stored in the Rucio catalogue. Rucio does not index the experiment-specific metadata in the files, however, they can be added through Rucio's generic metadata explicitly.

Files always have an availability status attribute, namely {\em available}, {\em lost}, or {\em deleted}. If at least replica exists on storage, a file is in state {\em available}. A file is in state {\em lost} if there are no replicas on storage, while at the same time at least one replication rule exists for the file. A file is in state {\em deleted} if no replicas of the file exist anymore. New files enter the system usually by registering first the file, then registering the replica, then actually uploading the file to storage, and finally placing a replication rule on the file to secure the replica. The availability attribute is thus a derived attribute from the contents of the Rucio replica catalog.

Users can set a {\em suppression} flag for a DID. It indicates that the owner of the scope no longer needs the name to be present in the scope. Files that are suppressed do not show up in search and list operations on the scope. This flag can be ignored when explicitly listing contents of collections with a deep check.

The collection status is reflected by a set of attributes, most importantly {\em open}, {\em monotonic}, and {\em complete}. If a collection is open then content can be added to it. Collections are created open, and once closed they cannot be opened again. Datasets from which files have been lost can be repaired when replacement files are available, even if they are closed, but this is an administrative action not generally available to regular users. If the monotonic attribute is set, content cannot be removed from an open collection. Collections are created non-monotonic by default. Once set to monotonic, this cannot be reversed. This is especially useful for datasets that follow a timed process. A collection where all files have replicas available is complete. Any dataset which contains files without replicas is incomplete. This is a derived attribute from the replica catalog.

Finally, Rucio also supports archives, such as compressed ZIP files. The contents of the archive files can be registered as constituents, and when resolving the necessary locations of the constituents, the appropriate archive files will be used instead. Some protocols can support transparent usage of archive contents, such as ROOT with ZIP files. In that case Rucio automatically translates the respective calls into their protocol specific direct access format of the constituent.

\subsection{Accounts}

\begin{figure}[t]
  \centering
  \includegraphics[width=0.69\linewidth]{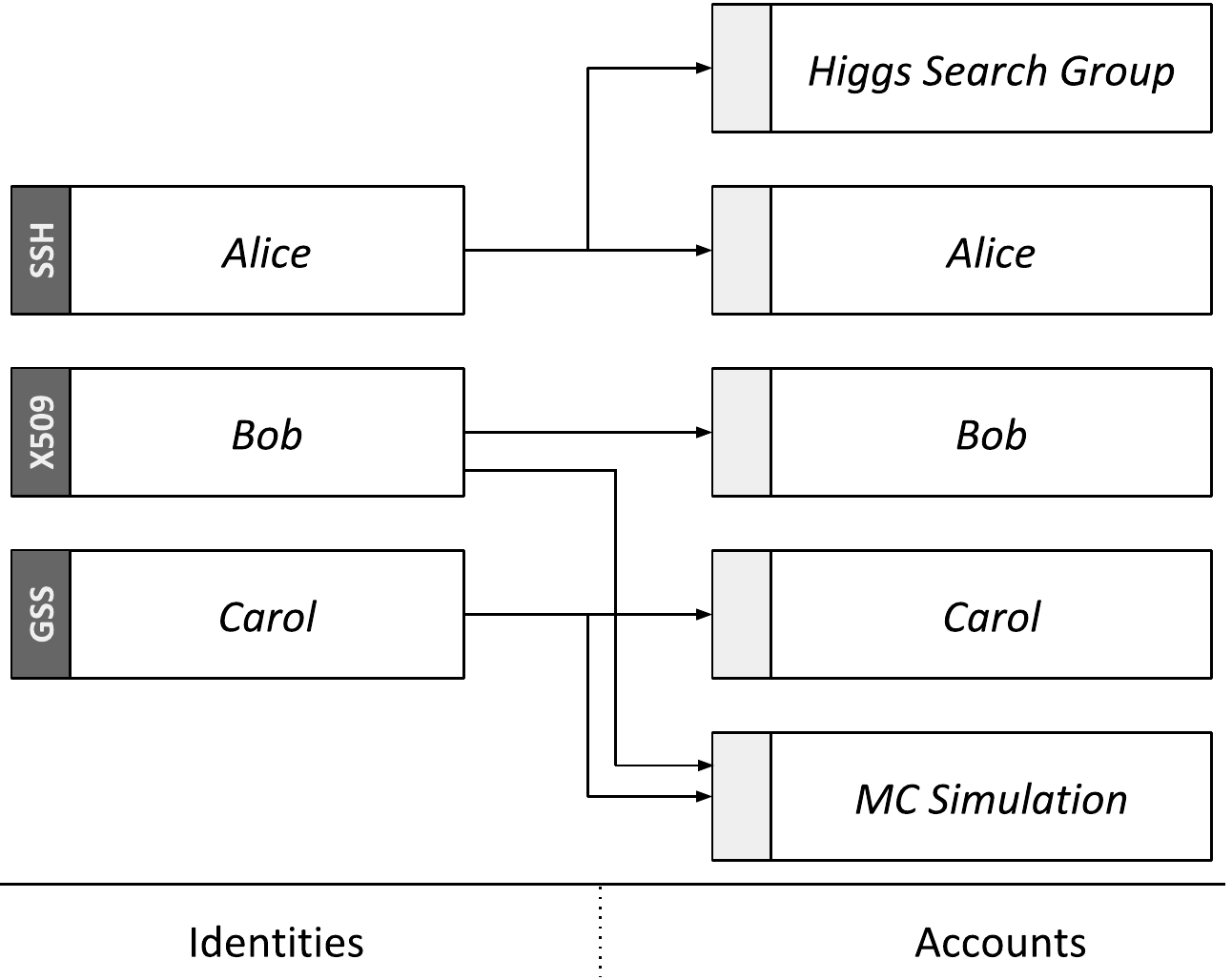}
  \caption{Each identity based on an authentication type can be mapped to one or more Rucio accounts and vice versa.}
  \label{fig:accounts}
\end{figure}

Each client that wants to interact with Rucio needs an account. An account can represent individual users such as Alice, a group of users such as the Higgs Search Group, or even organized activities like Monte Carlo simulation. A client can use different identities, sometimes called credentials, like X.509 certificates, username and password, SSH public key cryptography, or Kerberos tokens to authenticate to one or more accounts in a many-to-many relationship as shown in Figure \ref{fig:accounts}. The Rucio authentication system checks if the provided identities are authorized to act as the requested Rucio account. More details can be found in Section \ref{sec:auth}.

Each account has an associated scope in the global namespace, similar to a UNIX home directory. By authenticating to a different account, the client acquires access to different scopes. In the default configuration all data is readable by all accounts, even from private account scopes, but modification privileges are restricted. Privileged accounts can circumvent this restriction and modify the data across all scopes. These access permissions can be programmatically specified in the configuration to meet the needs of the organization using Rucio.

By default accounts have read access to all scopes and write access only to their own scope. This allows free sharing of data within a collaboration. Privileged accounts have write access to multiple scopes, for example, a workload management system is allowed to write into collaboration or user scopes as necessary.

If available, Rucio can retrieve data from external account systems, such as LDAP or VOMS, for automated account management.

\subsection{Storage}
\label{sec:storage_abstraction}

Rucio associates actual locations of the DIDs with Rucio Storage Elements (RSEs). These file locations are commonly called {\em replicas}. A single RSE represents the minimal unit of globally addressable storage and holds the description of all attributes necessary to access the storage space, such as hostname, port, protocol, and local file system path. RSEs can be extended with arbitrary key-value pairs to help create virtual spaces, allowing heuristics like {\em all tape storage in Asia}. Rucio allows permissions and quotas to be set for accounts independent from RSE settings. This allows flexible use of the available storage. No software services are needed at any of the data centers providing storage as RSE configurations are defined in Rucio.

File DIDs eventually point to the locations of the replicas. Each location is the physical representation of the file, i.e., bytes on storage. There can be files with zero replicas; these files would be denoted by the availability {\em deleted} and thus only exist in the namespace. For existing files on storage, these can be registered as-is directly into the Rucio catalog and will retain their full path information as given by the client. When uploading new data there are two possibilities, leave the decision of the path on storage to Rucio as automatically managed storage namespace, also known as {\em deterministic RSE}, or alternatively continue to provide full paths on the storage to the file, also known as {\em non-deterministic RSE}. Automatically managed storage namespace is programmatic and based on functions, e.g., using customizable hashes or regular expressions as detailed in Section \ref{sec:rmat}. Deterministic RSEs offer the advantage of access path calculation without contacting the Rucio file catalog. On the other hand, non-deterministic RSEs offer more flexibility in placing the files on storage.

Rucio is transport protocol agnostic, meaning that the RSEs can accept transfers via multiple protocols. As the RSEs are configured independently, the distribution of protocols can be heterogeneous and even depend on the location of the client accessing the data. For example, clients at the data center which try to access any of the local RSEs can use optimized transport using the {\em ROOT} protocol, whereas all clients from outside the data center could be served via HTTP/WebDAV. Protocol priority for read, write, deletion, and third party copy operations, as well as their fallbacks, are also supported, including proxy addresses.

Some commercial cloud providers use cryptographic signatures to control access to their storage paths. Currently this is supported for the Google Cloud Platform~\citep{gcp}, and is transparent to Rucio users. Elevation of privileges is automatically handled, and access control is regulated to ensure billing constraints. For S3-style cloud storage, either private or commercial, the distribution of pre-shared access credentials is also supported and directly associated with the RSE.

All connected RSEs have the notion of distance. This is not necessary geographical, but can be derived values, e.g., in ATLAS higher network throughput represents closer distance and is updated periodically and automatically. Functional distance is always a non zero value with increasing integer steps, and zero distance indicates no connection between RSEs. Most importantly, distance influences the sorting of files when considering sources for transfers. Practically, periodic re-evaluation of the collected average throughput of file transfers between two RSEs helps to dynamically adjust and update the distances to reflect the global state of the network and eventually improve source selection.

Some RSEs might allow data transfers, replica creation, and replica deletions outside of Rucio control. Such RSEs are considered volatile in Rucio. An example is a caching service which autonomously removes files based on high and low watermarks. Such RSEs are not presumed to guarantee data availability by Rucio at the time of access. It is then necessary that the caching service updates the Rucio namespace with timely location information by calling the appropriate API. Should the caching service fail to update the namespace appropriately, and a client thus cannot download or transfer a purported replica, then the replica will be flagged as suspicious and will be removed from the namespace. This could cause delays due to the retries, in case larger portions of the RSEs namespace are inconsistent.

\subsection{Subscriptions and replication rules}

The main mechanism for dataflow policies in Rucio are standing \textit{subscriptions}. Subscriptions exist to make data placement requests for future incoming DIDs, e.g., to automatically direct output of a scientific instrument to a particular data center. Subscriptions are specified by defining a metadata filter on matching DIDs, for example, all RAW data coming from the detector, and will automatically create the necessary replication rules, such as rules for tape archives in other countries. After the creation of a DID its metadata is matched with the filter of all subscriptions and for all positive matches the defined replication rules are then created on behalf of the account.

The actual replica management is then based on {\em replication rules} which are defined on the DIDs. A replication rule is a logical abstraction which defines the minimum number of replicas to be available on a list of RSEs. A replication rule affects the replication of all constituent files of this DID continuously. Thus when files are added or removed from a dataset, the replication rule also reflects these changes. Each replication rule is owned by an account. The amount of data in bytes covered by their respective replication rules is used to calculate the space occupancy of the account on an RSE.

Replication rules serve two purposes: to request the transfer of data to an RSE and to protect this data from deletion. As long as data is protected by a replication rule, i.e., a user expresses interest by placing a rule, then the replica cannot be deleted. Multiple accounts can own replication rules for the same DID on an RSE. In this case, the files are shared with only one physical copy, but the replicas are logically protected by multiple replication rules and are only eligible for deletion once all rules are removed.

A replication rule requires a minimum of four parameters to be created: the DID it affects, the RSE expression representing a list of RSEs where replicas can be placed, the number of copies of each file to be created, and the lifetime of the replication rule after which it is removed automatically. {\em RSE expressions} are described by a set complete language defined by a formal grammar to select RSEs. An attribute match of the grammar always results in a set of RSEs, which could also be empty. More detailed information of the language can be found in a dedicated article~\citep{replication_rules}.

\begin{figure*}[t]
  \includegraphics[width=\linewidth]{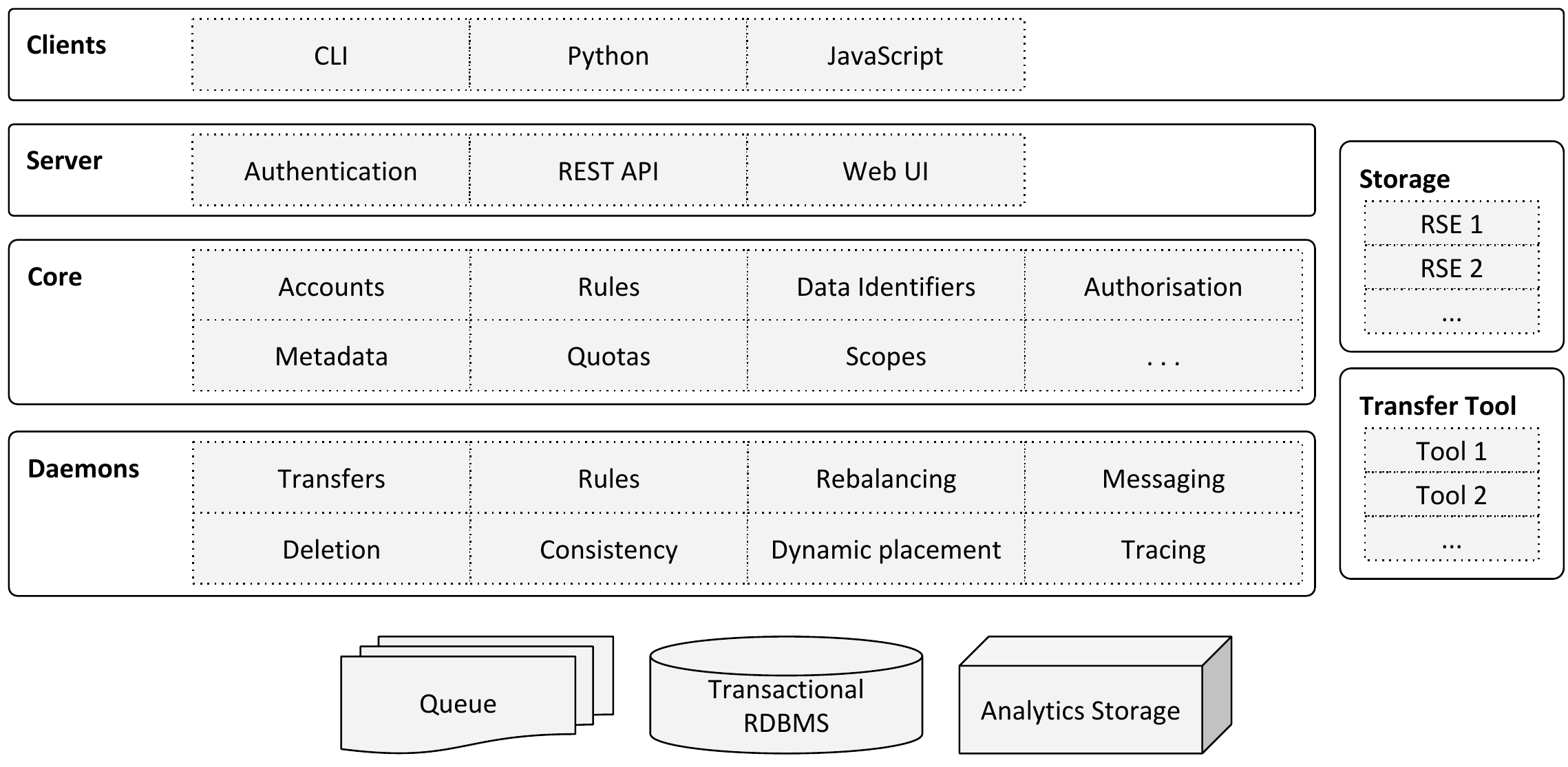}
  \caption{High-level Rucio component overview detailing the four layers: clients, server, core, and daemons. Storage and transfer tools are transparently integrated. The underlying support systems consist of queueing systems, transactional relational database management systems (RDBMS), and non-relational analytics storage.}
  \label{fig:rucio_architecture}
\end{figure*}

An RSE expression always evaluates into a list of RSEs. For example, the expression\\ "\texttt{tier=2\&(country=FR|country=DE)}" is equivalent to the set of all Tier-2s intersected with the set of all French and German RSEs. If the user defines an RSE expression with more RSEs than copies requested, it is up to Rucio to select where the data is being placed. Rucio primarily tries to minimize the amount of transfers created, thus it prioritizes RSEs where data is partially already available. Otherwise RSEs are selected randomly unless the \textit{weight} parameter of the rule is used, which allows the user to affect the distribution of data within the rule. The system internal bookkeeping of these selection decisions are called replica locks as they lock a replica on a certain RSE. A lock is always associated with a replication rule and once the placement decision has been made it will not be re-evaluated at a later point in time. This is to prevent the system from re-shuffling data continuously. Replica locks cannot be manipulated by users directly - their existence is always a result of an interaction with a replication rule. Examples of replication rules include
\begin{itemize}
	\item 2 copies of user.alice:myanalysis at \texttt{country=US} with 48 hours of lifetime
    \item 1 copy of user.bob:myoutput at \texttt{CERN} until January
    \item 1 copy of user.carol:testdata at\\ \texttt{country=DE\&type=tape} with no lifetime
\end{itemize}

When requesting the replication rule Rucio validates the available quota, evaluates the RSEs based on existing data, creates transfer requests if the data is not available at the specified RSEs, and creates the replica locks to prevent the data from being deleted. Until the removal or expiration of the rule the replica locks will prevent this data from being deleted. Notifications are always provided for state changes of rules and their transfer requests. These notifications are primarily useful to other systems for synchronisation purposes, e.g., notifying a workflow management system that a dataset has finished transferring.

There is no possibility of having conflicting rules, since the evaluation of rules always cause idempotent or additive results, i.e., either to keep the number of replicas as-is, or to create more replicas. It is not possible to restrict or limit other rules which could cause conflicting situations.

Finally, the measure of how much storage an account has used is derived from its replication rules. This can be controlled with quotas, which are policy limits which Rucio enforces on accounts. The accounts are only charged for the files they actively set replication rules on. The accounting is thus based on the replicas an account requested, not on the actual amount of physical replicas on storage. Thus if two different accounts set a replication rule for the same file on the same storage both accounts are charged for this file, although there is only one physical copy of it. The quotas can be configured globally and individually, and in case of overflows, can be approved or rejected by administrators.

\section{Architecture}
\label{sec:architecture}

\subsection{Overview}

Rucio, as shown in Figure \ref{fig:rucio_architecture}, is based on a distributed architecture and can be decomposed into four layers:
\begin{enumerate}
    \item the {\em clients} layer, such as the command line clients (CLI), Python clients, and the JavaScript-based web user interface,
    \item the {\em server}, offering the authentication, a common API for interaction with the clients and other external applications, and the WebUI,
    \item the {\em core} which represents the abstraction of all Rucio concepts, and
    \item the {\em daemons} taking care of the continuous and asynchronous work flows in the background.
\end{enumerate}
Next to these four main layers there are the {\em storage} resources and {\em transfer tools}, as well as the underlying persistence layer, represented by the different {\em queuing} systems, {\em transactional} relational databases, and {\em analytics storage} on non-relational databases.

\subsection{Clients layer}

The REST~\citep{rest} interface is the main entry-point to interact with Rucio. The client API simplifies the usage of the REST interface for Python environments. It implements the authentication, the correct use of already available authentication tokens, and Python wrappers for most of the REST interface commands. Furthermore there are different helper functions which encapsulate a set of API calls to implement more advanced functionality, e.g., downloading and uploading of files.

The functionality of the client API is organized into different classes, each one a subclass of {\em BaseClient}. The BaseClient provides the most important client information, e.g., the authentication token, the request session object, and account details. This class also implements the different authentication methods. There also exists a generic {\em Client} class which collects all client API calls into a single importable module for convenience. This allows the calling of all wrapped REST interface commands using a single object. The authentication is done directly by the BaseClient when creating any client class itself.

Rucio also comes with a variety of command line tools, e.g., {\em bin/rucio} and {\em bin/rucio-admin}. The first one provides all basic commands to interact with Rucio for non-administrative users. This includes listing DIDs, getting attributes or metadata, organizing DIDs, or downloading replicas. The other command line tool allows the execution of administrative commands such as adding newly available RSEs, or changing configuration attributes.

\subsection{Server and core layers}

The {\em server} is a passive component listening to incoming queries and forwarding them to the {\em core}. The core is the representation of the global system state. Incoming REST calls are received by a web server, such as Apache, and relayed to a WSGI~\citep{wsgi} container which executes the matching Rucio function in the core to update the system state. Any result of the function is streamed back to the WSGI container and Apache delivers it as an HTTP response object to the client. For more complex requests, such as the creation of large replication rules, which result in a large amount of transfers, the server accepts and confirms the client request, however the actual execution is done asynchronously by a daemon. This principle is commonly applied to ensure lower server utilization for fast response times and to be able to optimize the execution of large workloads in the background. In general, the servers do not directly interact with RSEs as these interactions are exclusively done by the daemons.

\subsection{Daemons layer}

The {\em daemons} are continuously running active components that asynchronously orchestrate the collaborative work of the entire system. Not all daemons are mandatory, some are optional such as the consistency or external messaging daemon. The daemons use a heartbeat system for workload partitioning and automatic failover. This principle enables automatic redistribution of the workload in case of a daemon crashing resulting in a lost heartbeat, but also to redistribute work when more daemons are started. Examples include: transfer daemons, which use a tool to submit queued transfer requests to the relevant service; and rule evaluators, which automatically re-evaluate replication rules which are  stuck due to repeated transfer errors.

\subsection{Storage and transfer tool layers}

The {\em storage} layer is responsible for the interactions with different grid middleware tools and storage systems. As explained in Section \ref{sec:storage_abstraction} an RSE is not a software run in a data center but only the abstraction of storage protocols, priorities, and attributes of a storage system, and can be configured dynamically and centrally. This abstraction effectively hides the complexity of distributed storage infrastructures and combines them in one interface to be used by all Rucio components. 

The {\em transfer tool} is an interface definition which must be implemented for each transfer service that Rucio supports. The interface enables Rucio daemons to submit, query, and cancel transfers generically and independently from the actual transfer service being used.

\subsection{Persistence layer}

The lowest layer is the persistence layer, or catalog, and it keeps all the data as well as the application states for all daemons. It requires a transactional database. Rucio uses SQLAlchemy~\citep{sqlalchemy} as its object relational mapper and supports multiple transactional relational database management systems (RDBMS) such as SQLite, MySQL, PostgreSQL, or Oracle. Upgrading the database schema is done via Alembic~\citep{alembic}, which emits the necessary SQL based on the updated relational mapping. Since downgrading is also supported the database schema can be restored to its previous state. In total, there are more than 40 tables representing the complete functionality of Rucio under full version control.

The catalog description is handled in Python natively and is used throughout the codebase instead of custom SQL statements. Several base classes have been developed that help with operational tasks, such as storing of deleted rows in historical tables, custom data-types, or automatically checking constraints. On the ATLAS Oracle instance some additional database internal helpers have been deployed, such as faster clean-up of sessions where clients had a timeout, or PL/SQL to help with fast calculation of table contents for operational statistics. The functionality of these scripts is also available via daemons, in case a Rucio instance is deployed without Oracle.

Interaction with the database has been highly optimized to eliminate row lock contention and deadlocks. Also targeted indexes on most tables have been added to improve database interaction throughput. In the ATLAS instance, selected tables have also been initialized with index-oriented physical layouts to reduce tablespace. Especially noteworthy is the work sharding across all instances of the daemons, where locking would be detrimental to SQL statement throughput. For Rucio, the selection of work per daemon is based on a hashing algorithm on a set of attributes of the work requests. All daemons of the same type select on the hashes to guarantee among each other not to work on the same requests. This works across all supported databases and allows lock-free parallelism per daemon type.

Helper scripts automatically extract table contents for storage in Hadoop~\citep{hadoop} for long-term backup of table contents, complex reports generation for annual reports, and off-site access for intensive clients. More details are discussed in Section \ref{sec:monitoring}.

\section{Functionality details}
\label{sec:functionality}

In this section, we describe how the most important data management functionalities of Rucio are represented based on the concepts and the architecture, i.e., the authentication and authorization, the replica management and transfers, data deletion, data consistency and recovery, messaging, and monitoring. Where necessary, we also highlight some implementation details.

\subsection{Authentication and authorization\label{sec:auth}}

Rucio supports several types of authentication: username and password, X.509 certificates with and without proxies, GSSAPI Kerberos tickets, as well as SSH-RSA public key exchange. Each successful authentication generates a short-lived authentication token, the {\em X-Rucio-Auth-Token}, which can be used for an infinite number of operations until the token expires. The token contains a set of identifying information, plus a cryptographically secure component. The token is cached locally on the client side and is secured with POSIX permissions based on the calling user.

Each subsequent operation against any of the REST servers needs the valid {\em X-Rucio-Auth-Token} set in its HTTP header. If the token has expired then the request is denied with an appropriate HTTP error code.

The implementations of the username/password and the SSH RSA public key exchange are native. X.509 authentication is performed via the GridSite library~\citep{gridsite}, and GSSAPI Kerberos authentication is performed via the ModAuthKerb library~\citep{modauthkerb}. Both serve as loadable modules inside the Apache HTTP server.

Authorization for specific functionality is also configurable and customizable. Each client-facing operation, such as listing datasets or deleting files, is validated through a permission function which can limit the allowed Rucio accounts. Every instance of Rucio can host different sets of permissions and can thus be customized to the access policies of each experiment.

\subsection{Replica management and transfers\label{sec:rmat}}

There are two workflows in Rucio which physically place data on storage: when a replica is uploaded via a client and when a replica is created by a transfer to satisfy a replication rule. In both workflows Rucio has to generate the physical path of the replica on storage. The system offers two paradigms to generate these file paths: {\em deterministic} and {\em non-deterministic}. A deterministically created path can be generated solely knowing the scope and name of a DID, ignoring the hostname or other knowledge of the RSE. A non-deterministic path requires additional information describing the file, such as meta-data, the dataset the file belongs to, and more. Non-deterministic paths are useful when the backend storage system requires specific location properties, such as co-located paths on tape drive systems. Non-deterministic RSEs are also useful when there are storage areas which are populated by outside systems, like the ATLAS Tier-0 prompt reconstruction facility, and are then registered in Rucio for later distribution.

Rucio supports pluggable algorithms for both deterministic and non-deterministic path generation which are defined per RSE. The {\em hash deterministic algorithm} is an algorithm commonly used in Rucio. The algorithm uses a one-way hash based on the name of the file to create the directory where it will place the file. Due to the characteristics of hash functions the files are distributed evenly over the directories, which is beneficial for the majority of filesystems where storage performance degrades based on the number of files in a single directory. For replicas created by transfers the same algorithms are applied to generate the path.

As explained in Section \ref{sec:concepts} transfers in Rucio are always a consequence of the rule engine trying to satisfy a replication rule. This is the only means by which users can request transfers. The internal workflow for transfer request handling works as follows:

\begin{enumerate}
	\item During the creation of the replication rule, {\em transfer requests} are created which define the target destination RSEs of the file.
    \item The registered requests are continuously read by the {\em transfer-submitter} daemon, which ranks the available sources for each request, selects the matching protocols of source and destination storage based on protocol priorities, and submits transfers in bunches to the configured transfer tool, which abstracts the underlying transfer service of the infrastructure.
    \item The {\em transfer-poller} daemon continuously polls the transfer tool for successful and failed transfers. Additionally, the {\em transfer-receiver} daemon observes a message queue and listens for successful and failed transfers. Most transfers are checked by the transfer-receiver, as its passive workflow decreases the load on the transfer tool.
    \item The last step is the {\em transfer-finisher} daemon which reads the successful and failed transfer requests and updates the associated replication rules.
\end{enumerate}

For failed transfer requests the transfer-finisher will update the associated replication rule as {\em STUCK}. Stuck rules are continuously read by the {\em rule-repairer} which will either decide to submit a new transfer request for an alternative destination RSE or re-submit, after some delay, a transfer request for the same RSE.

\subsection{Data deletion \label{sec:deletion}}

Deletion is intrinsically linked to rule and replica lifetimes. At the end of the rule lifetime replicas become eligible for deletion. A daemon continuously sets timed markers on such expired entries, or alternatively when the last rule pointing to the data was removed. The deletion daemon will then select the marked and expired entries and actually delete them from storage. This can happen in two different modes: greedy and non-greedy. Greedy mode removes data as soon as it is marked, which maximizes the free space on storage. Non-greedy mode deletes the minimum amount of data required to fulfill new rules entering the system, and keeps the existing data around for caching purposes. All thresholds involved in this process are configurable per RSE. The selection of files to remove is automatically derived from their popularity as given through their access timestamps. These timestamps are created by Rucio clients downloading the file, either through manually downloads or automatic downloads of input files as part of the workflow execution. This means that even expired replicas can stay at an RSE if they are still used within a configurable period.

In general, the lifetime of the rules and replicas follow experiment policies. In ATLAS, the policy depends on whether the replica is primary or secondary data: The disk-resident replicas, called primary data, are required to fulfill the ATLAS computing model or to serve ongoing activities like reprocessing or production. The cache data, or secondary data, is deleted in a Least Recently Used (LRU) manner when space is needed. Files in this category are replicated with a limited lifetime to serve more unscheduled activities like user analysis when necessary to reduce overall response time and bandwidth usage. They will be deleted automatically after the lifetime period. As a safeguard, in ATLAS all rule removals are configured with a 24h delay to undo any potential changes. As an additional safeguard, all highly important data, such as detector data, are protected both by replication rules issued by the root account without lifetimes, as well as RSEs with the deletion operation disabled.

\subsection{Data consistency and recovery}

\begin{figure}
  \centering
  \includegraphics[width=\linewidth]{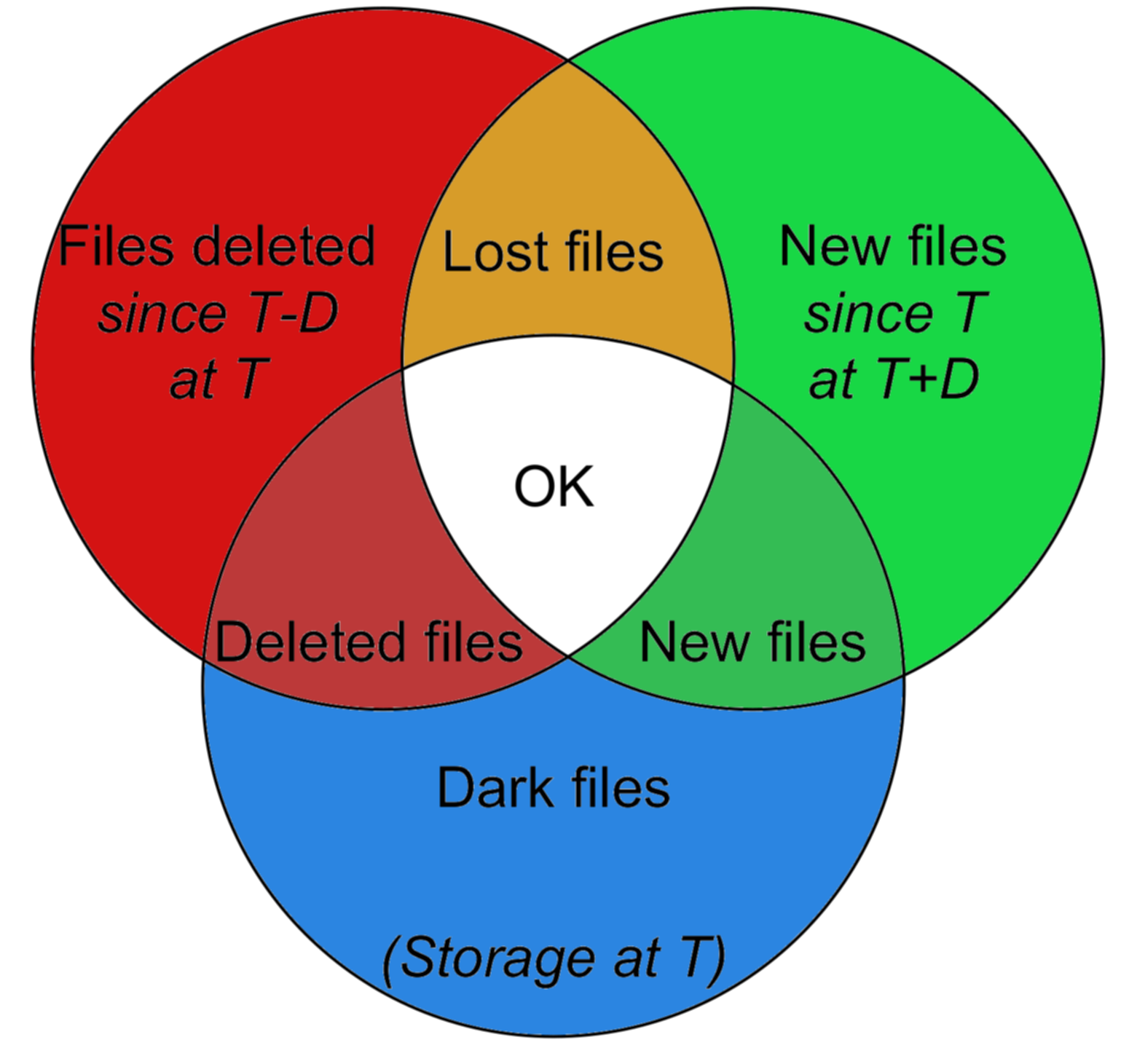}
  \caption{Consistency comparison diagram with their outcomes. The content of each catalog at a historical point in time (T-D), current point in time (T), and future point in time (T+D).}
  \label{fig:consistency}
\end{figure}

Different tools are available in Rucio to detect inconsistencies between the Rucio catalog and what is physically available on storage. One daemon is dedicated to identify {\em lost files}, i.e., files registered in the catalog but not present on storage, and {\em dark files}, i.e., files which exist on storage but are not registered in the catalog and must have been put on Rucio-managed storage areas through unsupported methods. It is important to remove dark files since the accounting and quota system depend on the correct state of the storage system with respect to the catalog contents. The actual comparison is done by evaluating the lists of files from storage and Rucio directly. The storage lists are provided periodically by the storage administrators and are accessible as plain text files at predefined places. Two comparisons are needed to check the contents of the storage lists from a given timestamp $T$, with the content of the Rucio catalog from an earlier time $T-D$ and a later time $T+D$. As such, the timestamp $T$ must always be historical, for example, one day in the past. Figure \ref{fig:consistency} describes the different categories of inconsistencies that can be detected using the three lists. The files found in all three lists are consistent both on storage and the catalog. The ones found on the two catalog lists but not on the storage list are lost files. The ones found in the storage list but not in the catalog are the dark files. All the other combination are transient, that is, new or deleted files which have yet to be signed off in their respective workflow. The dark files identified by this daemon are then deleted by the deletion daemon mentioned in Section \ref{sec:deletion}. The lost files are flagged with a special state for potential recovery.

Rucio also takes care of automatic data recovery in case of data loss or data corruption. Replicas can be marked as {\em bad} either by privileged accounts or by Rucio itself, when it detects that a replica has caused repeated failures, e.g., when used as a source replica or when downloads have checksum mismatches between the downloaded file and the checksum recorded in the Rucio catalog. A daemon identifies all bad replicas and recovers the data from another copy by injecting a transfer request if possible. In the case of the corrupted or lost replica being the last available copy of the file, the daemon takes care of removing the file from the dataset, updating the metadata, notifying external services, and informing the owner of the dataset about the lost data.

\subsection{Messaging}

Asynchronous communication with external systems is done via message queues. Rucio supports STOMP protocol compatible queuing services, e.g., ActiveMQ~\citep{activemq}, as well as email notifications. Every component can schedule messages for delivery to either STOMP or email providers. Each message consists of an event-type and a payload. The event-type can be used by queue listeners to filter for messages, such as {\em transfer-done} or {\em deletion-queued}. The payload is always schema-free JSON and can be arbitrary. Typically, the JSON payload consists of information about the operation and which data it acted upon, such as the protocol used for a transfer, or the time it took to delete a file.

These messages are typically used to asynchronously update external services with Rucio operations, e.g., workflow management systems are interested to listen for replication rule completeness events. Additionally, these events are used for monitoring and analytics as described in the next section.

\subsection{Monitoring and analytics \label{sec:monitoring}}

Rucio uses a variety of tools to monitor the different components of the system. Most importantly, there are three main monitoring systems: {\em internal}, {\em dataflow}, and {\em reporting}. These are split across a small number of different technologies due to different requirements on the storage and display of the monitoring data. The {\em internal} monitoring is used to follow the state of the system metrics, such as the size of queues or server response times. The monitoring for the {\em dataflow}, such as transfers and deletions, is used to check throughput or bandwidth and to identify problems due to user interactions, for example, trying to move Petabytes of data to a single data center. And lastly, regular {\em reports} and summaries are provided which are used for a variety of tasks, for example, site administrators use them to verify the usage of their storage, the consistency daemon uses them to check for storage problems, reports of the usage and access patterns of data are included in resource planning, and many more.

\begin{figure}[t]
  \includegraphics[width=\linewidth]{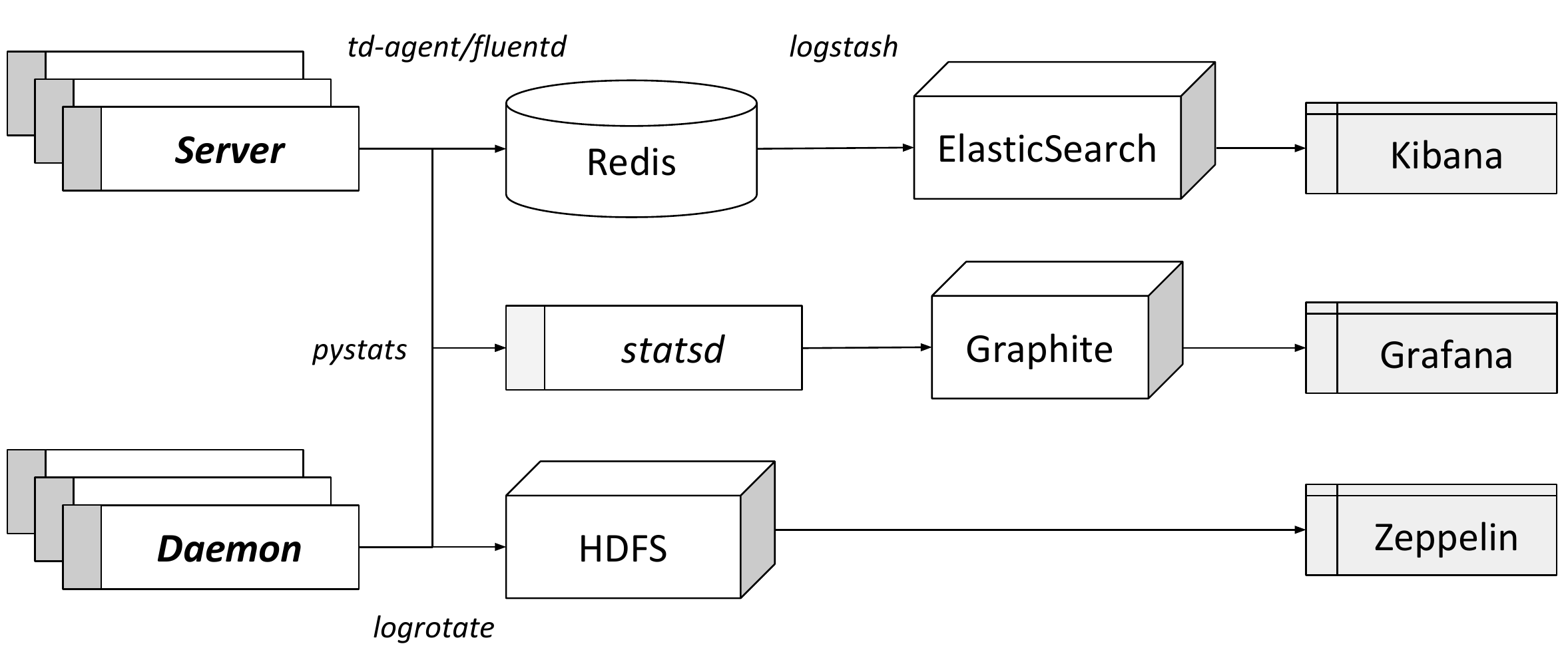}
  \caption{Overview of the internal system monitoring using pystats on the server and daemons to send metrics to a central statsd collector and from there to the Graphite database. Grafana is used to visualize the metrics. System logs are monitored using td-agent to send the logs from the server and daemon machines to a Redis buffer. From there Logstash collects them and writes them to Elasticsearch. Furthermore, daily {\em logrotate} jobs also write the logs to HDFS for backup, and can be examined via the Zeppelin web-based notebook}
  \label{fig:internal_monitoring_c}
\end{figure}

\begin{figure*}[t]
  \includegraphics[width=\linewidth]{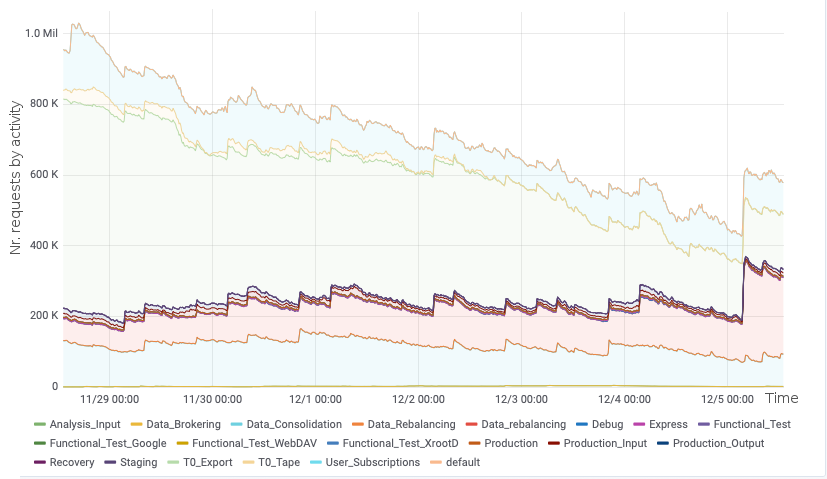}
  \caption{One example plot from the internal Grafana dashboard used to monitor the internals of the system. This plot shows the number of requests submitted to FTS split by activity over time.}
  \label{fig:grafana_example}
\end{figure*}

For the internal monitoring, pystats~\citep{pystats} is used directly in the core components of Rucio and in probes to report internal metrics. pystats is a Python client library for statsd~\citep{statsd} which is a network daemon that listens for statistics, such as counters and timers, and aggregates them to send them to a Graphite~\citep{graphite} server. An example of this is the reporting of queue sizes per activity for the transfer daemon. A probe regularly checks the database and sends the metrics using a counter to the central statsd server. From there they are aggregated and flushed every 10 seconds to Graphite. The overview of the architecture is shown in Figure \ref{fig:internal_monitoring_c}. For visualization Grafana is used. An example dashboard can be seen in Figure \ref{fig:grafana_example}.


Another part of the internal monitoring is based on the server and daemon logs. These logs are sent to Elasticsearch~\citep{elasticsearch} to be visualized with Kibana~\citep{kibana}. To do this a td-agent~\citep{tdagent} collector is running on each server and daemon node that reads the local log files and streams them to a central Redis~\citep{redis} data store. The data in Redis is only stored temporarily to act as a buffer. A Logstash~\citep{Logstash} daemon collects the logs from there to write them to Elasticsearch. Logstash is not just forwarding the data to Elasticsearch but also parses some of the data and adds additional information, e.g., it translates numerical values to a human-readable format. The logs are then written into different Elasticsearch indexes for server and daemons. Different dashboards are available to check the server API usage, the API errors, or the daemon activities in a similar style as the internal monitoring. Additionally there are daily {\em logrotate} jobs on each node to send the logs to HDFS~\citep{hdfs} for backup and long-term storage. In some cases, the logs on HDFS can be analysed via special web-based notebooks using Zeppelin~\citep{zeppelin}.

Another component associated with internal monitoring is transfer and deletion monitoring. This monitoring is mainly based on two other systems: the traces and the events. The traces are access information reported both by the ATLAS computing job execution environment, known as pilots, of the workload management system and the Rucio CLI tools. Every time a file has been used as input for a job, and therefore has been copied to any execution environment, a trace is created that is then sent to the central Rucio server via HTTP. The same is done for output files that are written back from the execution environment to the storage. Similarly, when a user downloads or uploads data from or to the storage using the Rucio CLI the same traces are sent. The Rucio servers forward these traces to an ActiveMQ topic from which it is distributed into different queues for several applications, one of which is then used by the monitoring framework. Next to the traces there are also Rucio events for deletion and transfers on storage themselves. The transfer daemon produces messages when a transfer is submitted, waiting, done, or failed which are then written to a database table. The same is done by the deletion daemon. From there the messaging daemons pick up the messages and send them to a topic on ActiveMQ.

\begin{figure}[t]
  \includegraphics[width=\linewidth]{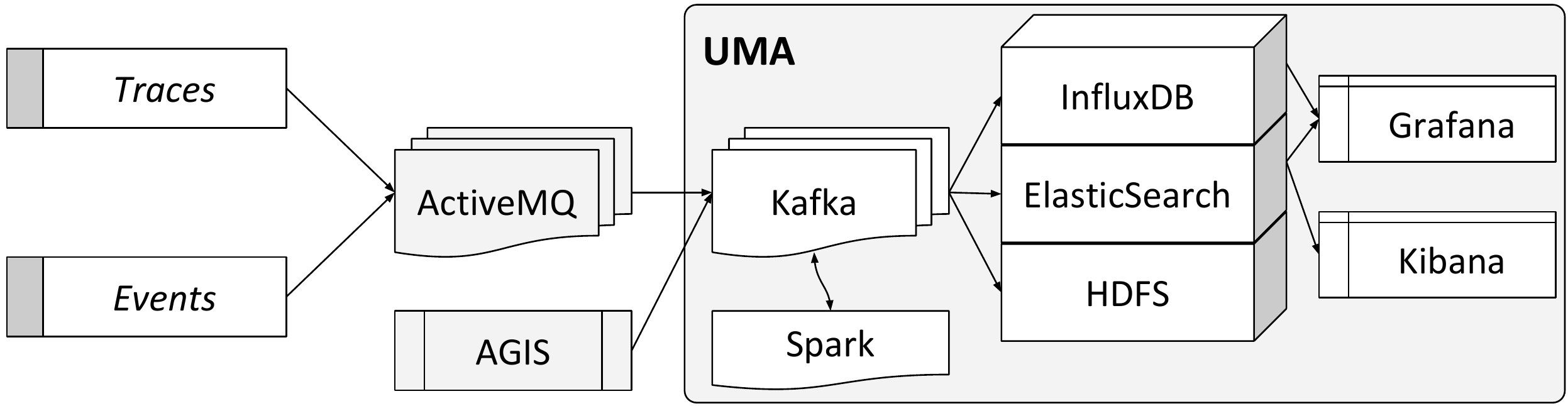}
  \caption{Overview of the monitoring architecture. The traces and events are sent to ActiveMQ from which they are forwarded to a Kafka processing queue. A continuously running Spark job aggregates and enriches the data and writes it back to Kafka. From there collectors write the data to HDFS for backup, Elasticsearch for detailed searches, and InfluxDB to be used with Grafana to produce the dashboards.}
  \label{fig:ddm_monitoring_arch}
\end{figure}

\begin{figure*}[t]
  \includegraphics[width=\linewidth]{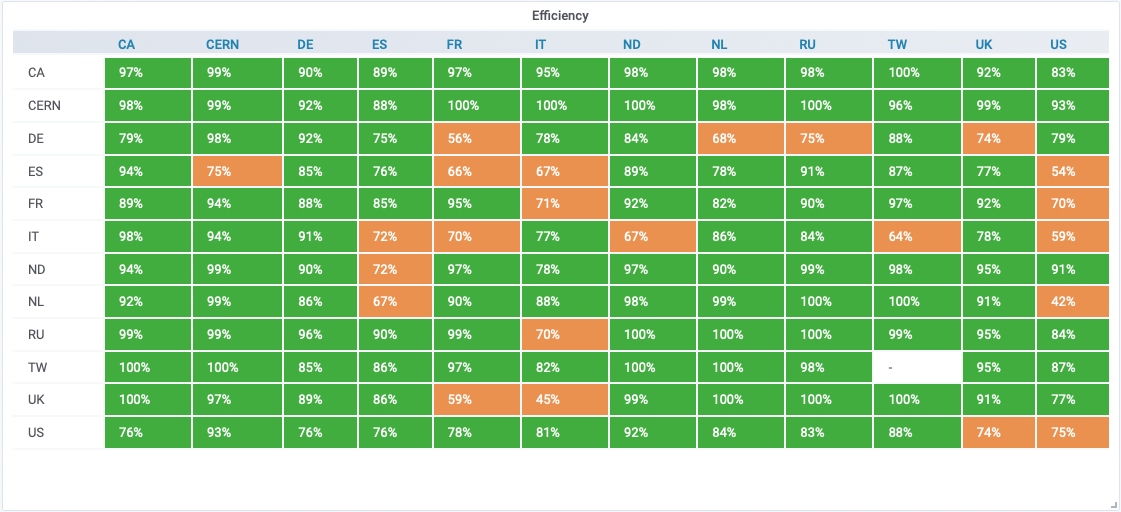}
  \caption{Matrix from the Grafana dashboard based on CERN IT UMA. It shows the efficiency of transfers between the source on top and the destination on the side, depicting geographical regions.}
  \label{fig:ddm_monitoring_vis}
\end{figure*}

The operational monitoring is provided by the Unified Monitoring Architecture (UMA)~\citep{uma} of the CERN IT monitoring team and is described in Figure \ref{fig:ddm_monitoring_arch}. The UMA is based on collectors that retrieve the traces and events from ActiveMQ and put them in Apache Kafka~\citep{kafka} and is enriched with topology information from the ATLAS Grid Information System (AGIS)~\citep{agis}, such as country or facility names. From there, continuous Spark~\citep{spark} jobs running on a Hadoop cluster get the data from Kafka, aggregate and further enrich the data, and write them back to a different topic in Kafka. Then the aggregated data is written to different storage systems: HDFS for long term backup, Elasticsearch for detailed searches, and InfluxDB for real-time monitoring. Everything is then used together in Grafana dashboards as shown in Figure \ref{fig:ddm_monitoring_vis}. There shifters, site-admins and operations can check the transfer efficiency, throughput, bandwidth and more and they also can drill down to find possible error reasons for failing operations.

The last important system used for monitoring is simple CSV lists produced on a regular basis. These lists are created using data imported to HDFS from different sources. Sqoop~\citep{sqoop} is used to import the important tables from the database, such as replicas, DIDs, dataset contents, or RSEs, and Flume is used to stream the traces directly from an ActiveMQ topic to HDFS. Then a set of daily and weekly Pig~\citep{pig} jobs are run on Hadoop to combine and process this data to create a variety of reports. They are provided as CSV files which can be read by users directly from Hadoop using Tomcat~\citep{tomcat} containers. The most important daily reports are the list of file replicas per RSE, which is used by the consistency daemon, lists of dataset locks per RSE used by site administrators to monitor site usage of their users, and a full list of all available datasets according to a specific pattern for the test system HammerCloud~\citep{hammercloud}. The weekly reports include lists of suspicious and lost files for site administrators, dataset access statistics based on traces for centrally managed storage areas, and a list of unused datasets that are used in reports for resource planning groups. Detailed storage accounting is available both as CSV lists and also in Elasticsearch for easy access for management and physics groups.

\section{Operational experience}
\label{sec:ops}

\subsection{Development workflow}

The current development workflow is the result of several years of experience and iterations with a large distributed development team. It is distributed under the Apache V2 license and thus free and open-source software. The development workflow relies on an agile development paradigm with time-based releases, and selected long-term support releases.

Since Rucio is a full-stack open-source project we rely on established tools in the open-source community, most notably GitHub~\citep{github} for version control management and Travis~\citep{travis} for automated testing.

Prior to any development, a traceable issue has to be created on GitHub describing the change, the planned modifications, the severity of the issue, and the affected components. This gives the entire developer community the chance to discuss the issue and point out possible implications. Each development gets classified into one of three categories: {\em feature}, {\em patch} or {\em hotfix}, which corresponds to the type of release in which the change will be included. Feature developments include database schema changes, new features, API changes, or any other larger enhancements. Patch developments include bugfixes or smaller enhancements to components. Hotfixes address a specific critical bug which require an immediate software release and are commonly done within the integration testing so the bug will not reach production. The release cycle is as follows: Patch releases (1.17.{\em NN}) are issued every two weeks; Feature releases (1.{\em NN}.00) are issued three to four times a year, mainly corresponding to LHC technical stops; Hotfix releases (1.17.3.post{\em NN}) are issued on-demand whenever necessary. Scripts are provided to help with this.

Modifications to the code are submitted as pull requests. These are merged into distinct branches, such that future feature developments do not impact the patch developments of the current feature branch. At the point of issuing a new feature release all future patch developments are based on this feature release. At the moment there is no long term support for previous feature releases, thus patches will not be issued for previously released versions. The release model is based on the requirements of the organizations currently using Rucio and can be evolved based on the future landscape of the Rucio community.

All pull requests are automatically tested by the Travis tool. Currently there are over 400 unit tests which are executed against several databases, such as Oracle, PostgreSQL, MySQL, and SQLite in Python 2.6, 2.7, 3.5, and 3.6 environments. We emphasize test-driven development, thus it is the responsibility of each developer to supply a good coverage of test cases for their developments. We require human review of all pull requests, which is open to the entire developer community. For a pull request to be merged it requires approval by at least one member of the core development team of Rucio. Merging is then done by the Rucio development lead who ensures the long term stability of the project and curates the tickets and release roadmaps.

\subsection{Deployment schema}

\begin{figure*}[t]
  \includegraphics[width=\linewidth]{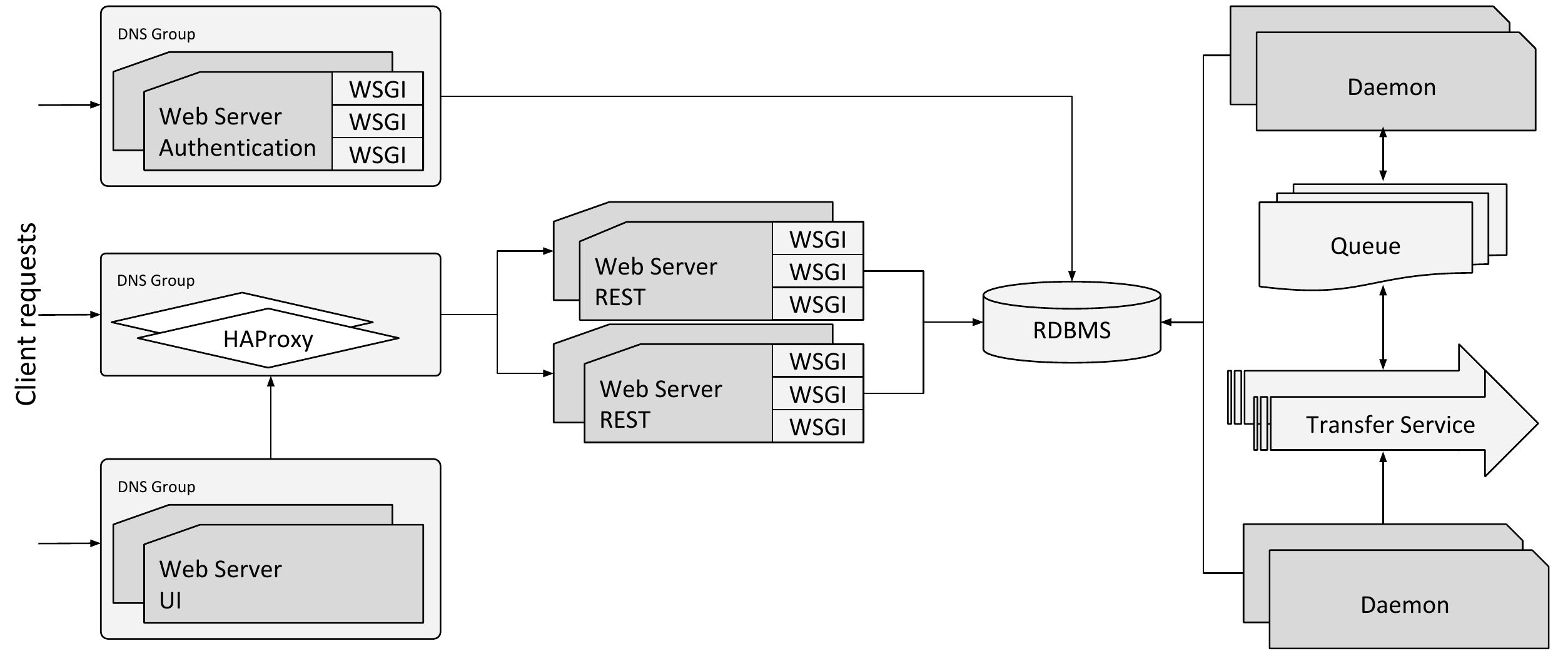}
  \caption{The deployment schema is able to accommodate multiple instances of each component for robustness and throughput.}
  \label{fig:deployment_schema}
\end{figure*}

Figure \ref{fig:deployment_schema} shows the recommended deployment schema. It allows robustness and horizontal scalability of the service by accommodating multiple instances of each service and daemon. This allows to theoretically infinitely scale the system, up to the point of the IO throughput of the underlying database. The actual volume of data to manage is thus irrelevant to the scaling of Rucio, only the number of entries in the catalog are affected by the database.

There is also the possibility to run a minimal Rucio system, either on bare metal or virtual machines, with good performance. This can already be accomplished by any off-the-shelf node with 4 cores and 8 GB of memory, a separate node for the database, as well as a separate node for FTS if third-party-copy functionality is needed. There is no installation of software needed at any of the participating data centers.

As shown in Figure \ref{fig:deployment_schema}, the clients operate on three endpoints, the authentication, the REST API, and the graphical web-based user interface. Each of these endpoints is in a domain name service (DNS) load-balanced group, which comprises the perimeter network of the data center. It is recommended to place a load-balancer, such as HAProxy~\citep{haproxy}, in the perimeter REST API, and keep the servers within the trusted networks. This way, clients of different kinds can benefit through custom load-balancing rules to point to selected to backend REST API servers. The authentication group is a separate DNS load-balanced group, both for separation of privileges as well as web-server authentication module encapsulation. The web-based UI itself is not within the trusted network, and communicates via asynchronous JavaScript with the REST API load-balancer. This allows us to treat web-based clients fairly when compared to programmatic access, and also reduces the risk of service stability attacks.

The web server itself spawns multiple instances and each instance controls multiple WSGI containers which execute the Python code. The default combination of Apache HTTP Server, {\em mod\_ssl}, {\em mod\_auth\_kerb}, and {\em mod\_gridsite} for authentication, as well as {\em mod\_wsgi} as the WSGI container have shown to be extremely reliable and efficient.

Almost every component has to interact with the central database. Modern sharding, partitioning, and hot swapping techniques for databases reduce the risk of service failure in case of database problems. For a very large database installation, dedicated expert effort is recommended. The system internal monitoring also provides detailed views on the database state, so early interventions on the database are easily possible by database administrators.

The daemons run inside the trusted network and interact with the database directly for performance reasons, they do not go through the REST API and are thus privileged. Each daemon can be instantiated multiple times in parallel, both for service robustness and horizontal scalability, just like the servers. Some daemons also interact with the message queue, both producing and consuming messages. STOMP-protocol compatible services such as ActiveMQ have shown to be stable and scalable. Other daemons interact with the underlying transfer services via the transfer tools. Redundant installations of the FTS system across the globe has shown to be stable and scalable as well.

\subsection{System performance}

Rucio's performance can be evaluated in several ways, most importantly, volume of managed data, interaction throughput, client response delay, database utilization, and node utilization.

\begin{figure}[t]
  \includegraphics[width=\linewidth]{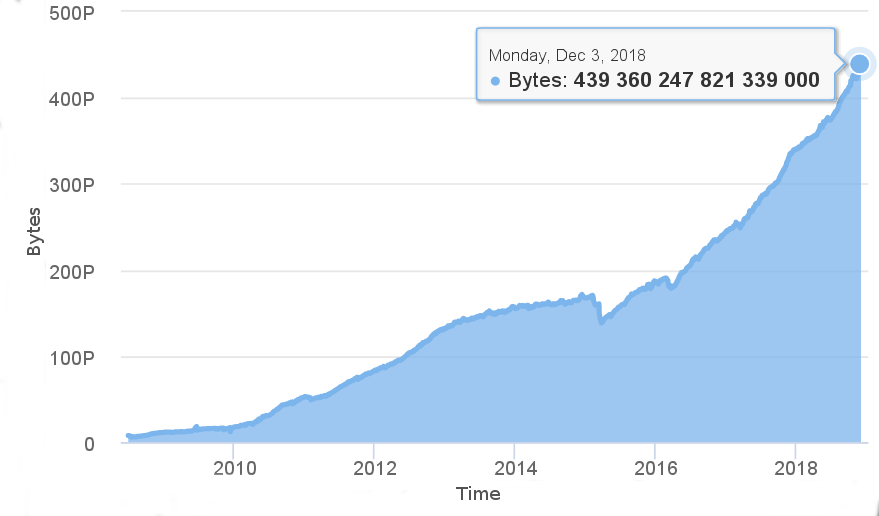}
  \caption{Total ATLAS volume managed by Rucio, approaching 450 Petabytes of data at the end of 2018.}
  \label{fig:atlas_volume}
\end{figure}

The largest Rucio deployment to date is for the ATLAS Experiment. The full deployment is hosted in the CERN data centre, including the database and the main FTS server, with additional FTS servers in the United States and the United Kingdom. As shown in Figure \ref{fig:atlas_volume}, the total volume of data approached 450 Petabytes by the end of 2018, with linear growth rates both during and between data taking periods. This data includes centrally produced experiment data, such as detector data and Monte-Carlo simulation, but also user data from individual data analysis groups or persons. At the end of 2018 the number of DIDs was 25 million containers, 13 million datasets, and 960 million unique files. The curious skew between containers and datasets is due to the use of containers for grouping of physics simulation, automatic data derivation and processing, and user analysis; the datasets themselves are mainly used as the unit of parallel workflow processing, thus having a comparatively small number of files per dataset. The number of RSEs is 860 and the number of replicas is 1.2 billion across all disk and tape storage. There is no discernible performance difference on Rucio catalog operations for files on disk when compared to files on tape.

\begin{figure}[t]
  \includegraphics[width=\linewidth]{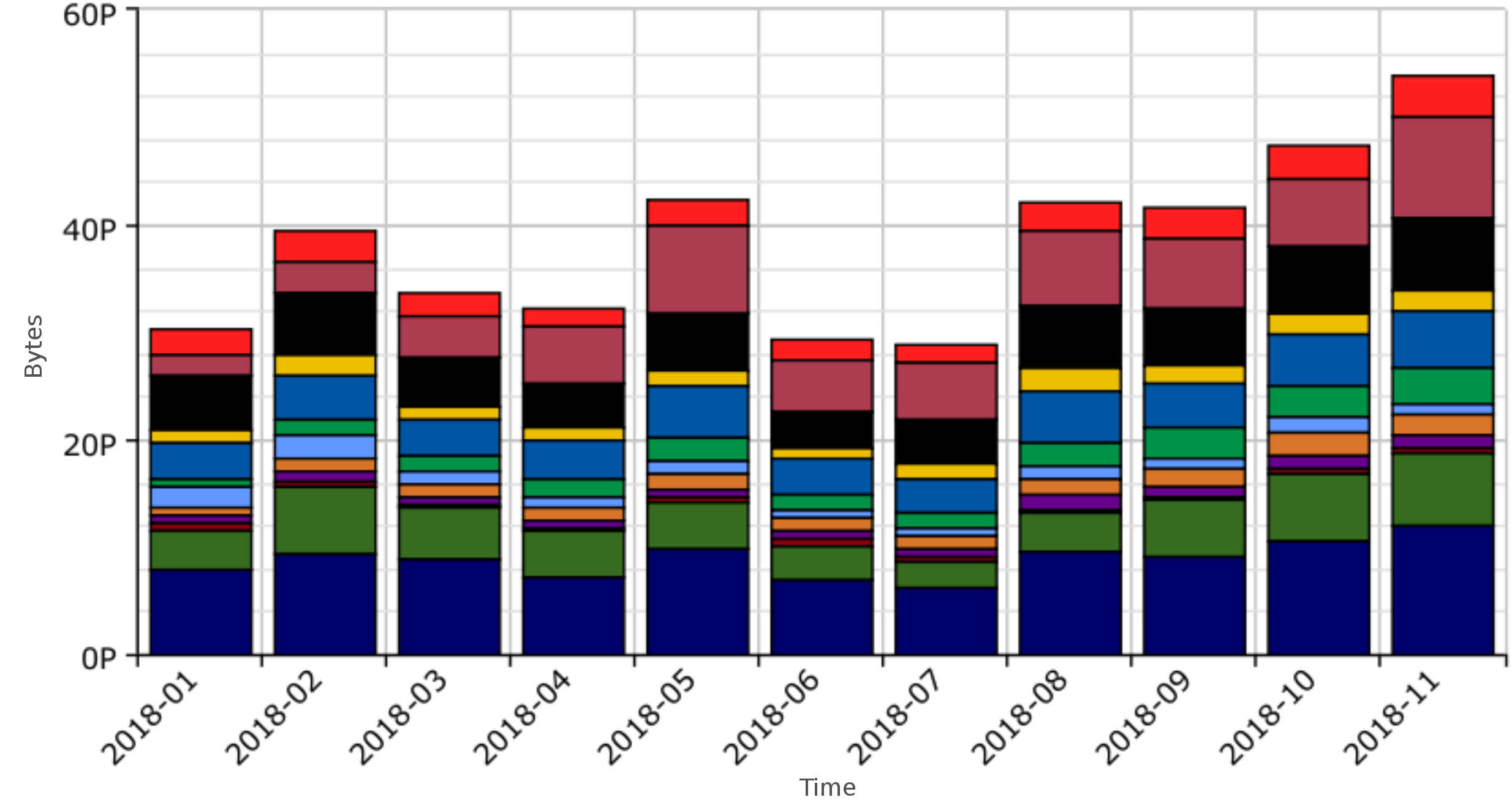}
  \caption{Total ATLAS volume transferred per month is consistently 30 Petabytes and above, reaching more than 50 Petabytes in November 2018. Colors denote different geographical regions.}
  \label{fig:atlas_transfer_rate}
\end{figure}

The data transfer and deletion rates are historically an indirect result of the computational needs of the experiment. As shown in Figure \ref{fig:atlas_transfer_rate}, ATLAS generally transferred at least 30 Petabytes of data per month in 2018, peaking at a record 55 Petabytes in November. The workload is quite regular both in the long term and short term, and there are few bursts with the exception of weeks leading up to physics conferences. On average 50 to 70 million files are transferred between data centres per month, with a transfer failure rate of roughly 10 million per month mostly due to storage and network configuration problems. These transfer failures are automatically recovered by Rucio, and the users do not need to worry about them. The deletion rate is higher than the transfer rate with up to 100 million files commonly deleted per month, amounting to 30 Petabytes and more, with an error rate of 10 to 20 million per month. Again, these are mostly attributed to storage configuration problems related to authorisation. The large number of files to delete are mostly intermediate data products stemming from computational physics workflows. Tape recall is considerably lower than transfers. Per month, ATLAS recalled about 1 Petabyte with fewer than 1 million files and with less than 10 percent recall issues that required recall retries. This high percentage usually comes from very large requests, which eventually time out and have to be retried. Historically, tape is considered as a write-only archive, but in case large samples are needed for simulation or users require raw data these can be staged from tape efficiently.

In addition to the interactive users, essentially all computational jobs interact with the Rucio server to locate and register data. The global server interaction rate is averaging 250 Hz, with frequent spikes up to 400-500 Hz. Average response time as measured by HAProxy is less than 50ms, though streaming the content of the replies can extend the total connection duration up to multiple seconds; this does not block other clients though. This interaction rate causes only low utilization of the 15 nodes (4 core CPU, 8 GB RAM) hosting the Rucio REST API servers at 5-7\% CPU utilization each and no measurable IO-Wait. The service is thus almost one order of magnitude over-provisioned just to ensure robustness in case of catastrophic data center failures where single nodes might become unavailable.

The database usage is split into its CPU usage, its session handling, and its tablespace volume. In the CERN Oracle instance with 16 logical CPUs, Rucio core utilization is on average about 20\%. To keep up with the high interaction rate, session sharing is used which keeps the number of active sessions consistently below 20. Peak streaming content from the database through the Rucio servers to the clients can easily reach up to 1500 concurrent sessions. Physical read rate on the database disk is less than 100 Mbps at more than 1 million IOPS, corresponding to 3000 transactions per second. Frequent spikes to multiple hundreds of Mbps are possible though. Current total database volume is 3.7 TB, with consistent growth rates of around 1 Terabyte per year since 2015. A hot standby of the database is hosted outside of CERN in the Geneva city center.

Finally, the operation of the ATLAS instance is covered by the core development team with {\em DevOps}-style full stack responsibility. In practice however, the only time the instance is actively touched is to upgrade to newer Rucio releases, which is negligible effort. Actual operation of ATLAS data management is a dedicated effort by a single person, who follows up on larger experiment requests, for example, massive data transfer campaigns above user-allowed quotas, to discuss experiment-wide configurations with the physics groups, or to pinpoint problems with storage and network. There are no ATLAS data operations persons at the data centres. The administrators at the data centres are only responsible for the configuration and running of their storage, both in software and hardware.

\section{Advanced features}
\label{sec:advanced}

This section describes three advanced features that can be used separately or in conjunction. The effectiveness of these methods is entirely dependent on the actual dataflows of each experiment and should be carefully evaluated with real experiment workload. Some of them can even be counterproductive if important system metrics are not available, such as automated movement of data which is needed to cope with storage imbalance.

\subsection{Dynamic data placement}

On top of the rather static replication policies that make sure that the data is well distributed across the grid to make them available for analysis by users, dynamic data placement helps to exploit computing and storage resources by removing replicas of unpopular datasets and creating additional replicas of popular ones at different RSEs. New replicas are created if a threshold of queued jobs is exceeded, taking into account the available resources, dataset popularity and network metrics to make sure that the new replicas are created quickly. Especially the number of queued jobs is always specific to the actual computation workload of the experiment, therefore this requires an interaction with the experiment's workflow management system. The currently used algorithm concentrates on free space and network connectivity between sites, so it weighs each site based on those criteria to find a suitable storage endpoint but also ensures that it does not put too much stress on single RSEs.

The current configuration of the dynamic data placement tool constantly scans incoming user jobs and collects the input datasets. The placement algorithm runs for every dataset containing official Monte Carlo or detector data. First the algorithm checks if there has already been a replica created in the recent past. It then checks how many replicas already exist below a configurable threshold and the popularity of the dataset. The algorithm continues to check network metrics for links between RSEs having an existing replica and possible destination RSE, as well as metrics such as free space, bandwidth and queued files, or if other replicas have been recently created there. If a suitable RSE has been found, the algorithm creates a replication rule, which will then take care of the transfer. Finally, detailed information about the decision is written to Elasticsearch for further analysis by operators and infrastructure providers.

On average 60 percent of these newly created replicas were quickly used again by the workload management system, i.e., within two weeks. On a longer time scale, half of accessed datasets are accessed more than once, i.e., the algorithm successfully creates replicas that are popular for several months.

\subsection{Automated data rebalancing}

Data rebalancing is a very common workflow in distributed data management systems, historically carried out by human operators. Rucio offers an automated service for these rebalancing workflows. The service provides three separate modes of operation: automatic {\em background} rebalancing, RSE {\em decommissioning}, and {\em manual} rebalancing executed by a human.

The automatic background rebalancing mode aims to equalize the ratio of primary and secondary replicas on a set of RSEs. By means of an equalized ratio it is guaranteed that there is relatively the same amount of secondary replicas available for deletion to make space for new data. For example, for the ATLAS experiment the background rebalancing is active for all RSEs with larger storage capacities. In each iteration, the algorithm calculates the average ratio and moves data from the RSEs above ratio to RSEs below it. The selection criteria for the data can be modified, but older, unpopular data, with a long lifetime is preferred. Replication rules are selected whose RSE expression would not conflict with the new destination RSE. The service links the original replication rule with the newly created one and only allows the removal of the original rule once the data has been fully replicated. The maximum volume of data and files to be transferred per day can be configured, as well as the activity of the transfers to separate the data from more important transfer activities. This helps to avoid overload of the storage and network resources with rebalancing activities.

The decommissioning mode allows an operator to select an RSE for removal from the infrastructure. The task of migrating RSE data in order to be able to decommission it is very labor intensive and error prone, especially when unique data is located at the RSE. However, with this service it can be done very quickly and safely. In contrast to the background rebalancing mode, where only some data is selected to be moved, the decommissioning mode selects all data resident on the RSE and moves it to a different RSE, following the original RSE expression policies of the individual rules.

The manual rebalancing mode allows operators to move a certain volume of data away from an RSE, in case of storage shortages or other data distribution considerations. This can be triggered at any time. The internal workflow is similar to the background rebalancing mode, with the operator only needing to specify the amount of volume to be rebalanced from an RSE.

\subsection{Transfer time prediction}

A trace record is created for every single transfer managed by Rucio. Most importantly, this includes the selected source and destination, the file size, and a series of timestamps indicating milestones in the transfer life cycle. It is possible to apply large-scale statistical analysis techniques in order to characterize the time spent for every transfer in each of its life cycle stages and thus predict the characteristics of large-scale data movement to improve task scheduling, network and storage optimization due to better endpoint selection~\citep{wesley}.

Rucio supports extension modules which can access these internal instrumentation data. The {\em Transfer Time To Complete} ($T^3C$) extension explores the possibility to model the transfer characteristics, with the aim of providing reliable transfer time estimates to Rucio core and other clients. In the general case, when a user creates a new rule, Rucio will reply with an estimate of when the rule will be finished. This includes calculations across all potential file transfers necessary to satisfy the rule. The module allows use of simultaneous models and features the ability to easily compare their performance. This extension opens the possibility for interested students to develop new machine learning algorithms to model the system characteristics.

\section{Summary}
\label{sec:summary}

\subsection{Conclusions}

Effective management of large sets of data, both in terms of volume and namespace, has been shown to be an extremely difficult problem. Rucio provides a solution to this problem with demonstrated usability, performance, scalability, and robustness, allowing scientific collaborations to fully use their distributed heterogeneous storage resources. The design of policy-driven data management has proven to be an excellent choice, giving the users the possibility to express their needs without having to worry about how to actually achieve them, but also because it allows the system to optimize itself during runtime based on self-instrumentation. The modular and horizontally scalable architecture has also been shown to handle the load of the ATLAS Experiment, and allows for the possibility to improve components selectively without having to re-engineer the core of the system.
It is important to mention that this came out of operational experience over many years which helped to address all these issues. The monitoring of the system has been especially well-received by the users, giving them detailed insights into their data management workflows. Synchronization with external systems is also decoupled through API backwards compatibilities and asynchronous messaging, giving both Rucio and external systems the possibility to evolve separately. Finally, the integration with storage, network, and transfer follows clear interfaces but custom implementations, which allows Rucio to benefit from individual optimizations which are exposed by different providers and can be extended quickly.

\subsection{Outlook}

The Rucio community is also growing. The system is now in use by two additional collaborations in production, ASGC/AMS~\citep{AMS} and Xenon1T~\citep{Xenon1T}, currently being deployed into production by the CMS~\citep{CMS} and DUNE~\cite{DUNE} experiments, and and is being evaluated by many other collaborations also outside of the high energy physics field, such as LIGO~\citep{LIGO} and SKA\citep{SKA}.


The future development of Rucio follows a dual approach: support for the High Luminosity LHC data needs, as well as the development of features relevant to the non-LHC communities.

With the start of the High-Luminosity LHC (HL-LHC), the year 2026 will see a significant increase in the data rates due to the increase in the number of HL-LHC collisions, higher event triggering accept rates, and more data products in offline computing. This presents both a funding and technological challenge, and will require several research and development initiatives. Currently identified topics are focusing on smart content delivery, data staging, and caching. Rucio will take the role as the orchestrating component, and ensure the reliable and efficient communication among the participating components, such as Software Defined Networks (SDNs)~\citep{sdn}, caching services~\citep{caches}, and High Performance Computing (HPC) centers~\citep{hpc}. These new workflows and functionalities will require integration and development efforts without breaking the existing horizontal scalability of Rucio.

Support for non-LHC communities will be driven by their requirements, and are mainly coming from the neutrino and astronomy sciences. The major feature requests include the support for arbitrary and mutable metadata, which was recently implemented, flexible user data synchronization and sharing, a stand-alone file transfer service for non-grid-style storage systems using WebDAV or HTTP, interfacing with other workflow management systems such as HTCondor~\citep{htcondor} or DIRAC~\citep{dirac}, time-based embargoes of data for scientific publications, as well as connectors to research databases like Zenodo~\citep{zenodo} to link publications to their data.

\subsection{Acknowledgements}

This work was done as part of the distributed computing research and development program within the ATLAS Collaboration. We thank our ATLAS colleagues for their support. In particular we wish to acknowledge the contributions of the ATLAS Distributed Computing (ADC) team. We also thank former colleagues Miguel Branco, Pedro Salgado, and Florbela Viegas for their contributions to the Rucio predecessor system DQ2.

\noindent On behalf of all authors, the corresponding author states that there is no conflict of interest.
\bibliography{bib}

\end{document}